%% file: main.tex
\documentclass[conference]{IEEEtran}
\IEEEoverridecommandlockouts

\usepackage[T1]{fontenc}
\usepackage[utf8]{inputenc}
\usepackage{csquotes}

\usepackage{amsmath}
\usepackage{braket}
\usepackage{pifont}

\usepackage{graphicx}
\usepackage{pgf} 
\usepackage{tikz}
\usetikzlibrary{quantikz2}
\usetikzlibrary{arrows.meta, calc}

\usepackage{siunitx}
\usepackage{tabularray}
\UseTblrLibrary{booktabs}

\usepackage{parskip}
\usepackage{caption}
\usepackage{subcaption}

\usepackage{xcolor}
\usepackage{orcidlink}
\usepackage{hyperref}
\usepackage{url}            
\hypersetup{colorlinks=true,urlcolor=teal,
  linkcolor=teal,citecolor=teal} 
\usepackage{glossaries-extra}

\usepackage[inline]{enumitem}
\usepackage{xspace}
\usepackage{censor}

\definecolor{codebg}{HTML}{F6F8FA}      
\definecolor{coderule}{HTML}{E1E6EB}    
\definecolor{codeaccent}{HTML}{002D4C}  
\definecolor{codetext}{HTML}{1B2A38}    
\definecolor{codegutter}{HTML}{A9B4BF}  
\definecolor{codekeyword}{HTML}{0B5C8C} 
\definecolor{codestring}{HTML}{00786A}  
\definecolor{codecomment}{HTML}{7A8794}
\definecolor{codeapi}{HTML}{B26A00}     

\usepackage[varqu,varl]{zi4} 
\usepackage{listings}
\usepackage[most]{tcolorbox}

\lstdefinestyle{pythonstyle}{
  language=Python,
  basicstyle=\ttfamily\scriptsize\color{codetext},
  keywordstyle=\bfseries\color{codekeyword},
  stringstyle=\color{codestring},
  commentstyle=\itshape\color{codecomment},
  emphstyle=\color{codeapi},
  emph={Script,execute,draw,dagger,power,evolve,Hermitian,PulseGates,
    ParametrizedHamiltonian,DepolarizingChannel},
  numbers=left,
  numberstyle=\ttfamily\tiny\color{codegutter},
  numbersep=9pt,
  xleftmargin=17pt,
  breaklines=true,
  breakatwhitespace=true,
  postbreak=\mbox{\textcolor{codegutter}{$\hookrightarrow$}\space},
  columns=fullflexible,
  keepspaces=true,
  tabsize=4,
  showstringspaces=false,
  aboveskip=0pt,
  belowskip=0pt,
}

\tcolorboxenvironment{lstlisting}{
  enhanced jigsaw,
  colback=codebg,
  colframe=coderule,
  boxrule=0.4pt,
  arc=3pt,
  outer arc=3pt,
  boxsep=0pt,
  left=3pt, right=5pt, top=5pt, bottom=5pt,
  borderline west={1.6pt}{0pt}{codeaccent},
  borderline west={0.4pt}{17pt}{coderule},
  before skip=6pt, after skip=6pt,
}

\usepackage{microtype}      

\newboolean{anonymous}
\setboolean{anonymous}{false}

\usepackage[
  backend=biber,
  maxcitenames=3,
  minbibnames=3,
  mincitenames=1,
  maxbibnames=3,
  uniquelist=false,
  url=true,
  doi=true,
  eprint=true,
  isbn=false,
]{biblatex}
\AtEveryBibitem{\ifboolexpr{test {\iffieldundef{doi}} and test {\iffieldundef{eprint}}}{}{\clearfield{url}\clearfield{urlyear}\clearfield{urlmonth}\clearfield{urlday}}}

\input{acronyms.tex}
\input{commands.tex}

\begin{document}

\title{JAQSI: Just Another Quantum Simulator}

\author{
  \IEEEauthorblockN{%
    \blackout{Melvin Strobl}\IEEEauthorrefmark{2}\orcidlink{0000-0003-0229-9897},
    \blackout{Maja Franz}\IEEEauthorrefmark{4}\orcidlink{0000-0002-2801-7192},
    \blackout{Lukas Scheller}\IEEEauthorrefmark{2}\orcidlink{0009-0003-9156-7781},
    \blackout{Eileen Kuehn}\IEEEauthorrefmark{2}\orcidlink{0000-0002-8034-8837},
    \blackout{Wolfgang Mauerer}\IEEEauthorrefmark{4}\IEEEauthorrefmark{5}\orcidlink{0000-0002-9765-8313},
    \blackout{Achim Streit}\IEEEauthorrefmark{2}\orcidlink{0000-0002-5065-469X},
  }
  \IEEEauthorblockA{
    \IEEEauthorrefmark{2}
    \blackout{Karlsruhe Institute of Technology, Germany},
    \{\genpartemail{melvin.strobl}{kit.edu}, \genpartemail{lukas.scheller}{kit.edu}, \genpartemail{eileen.kuehn}{kit.edu}, \genpartemail{achim.streit}{kit.edu}\}@\blackout{kit.edu}
  }
  \IEEEauthorblockA{
    \IEEEauthorrefmark{4}
    \blackout{Technical University of Applied Science Regensburg, Germany},
    \{\genpartemail{maja.franz}{othr.de}, \genpartemail{wolfgang.mauerer}{othr.de}\}@\blackout{othr.de}
  }
  \IEEEauthorblockA{
    \IEEEauthorrefmark{5}
    \blackout{Siemens AG, Foundational Technology, Munich, Germany}
  }
}

\maketitle

\begin{abstract}
  Classical simulation supports the development and validation of quantum algorithms through full-state inspection, controlled noise models and backpropagation, albeit for small quantum systems.
  Existing frameworks offer gate-level execution, differentiable pulse-level dynamics or optimised hardware execution.
  Just another quantum simulator (\jaqsi) combines gate- and pulse-level simulation in one differentiable circuit model built upon JAX.
  Circuits are represented as pure functions of numerical inputs which support vectorised batches and \glsxtrshort{jit} compilation.
  Within a single circuit, matrix-defined operations, gates synthesised from Hamiltonian time evolution and noise channels can be composed freely.
  For a fast simulation, gate and channel fusion reduces passes over the state, while adjoint differentiation computes expectation-value gradients with a constant number of statevectors.
  At pulse level, \jaqsi reduces evolution under single-term Hamiltonians to a scalar pulse-area integral and reuses cached execution plans and compiled solvers.
  In a benchmark we compare \jaqsi with PennyLane, Qiskit, Qibo, Qulacs, QuTiP and dynamiqs.
  At gate level, we observe performance either on par or an order of magnitude faster across statevector, density, noise and gradient simulation than the compared simulators up to \num{12} qubits.
  Beyond \num{18} qubits, \jaqsi is on par with PennyLane Lightning in both statevector and gradient simulation.
  In the statevector and noisy density-matrix pulse benchmarks, it is at least \num{5} times faster than the other simulators, even with its closed-form solve disabled.
  Disabling the latter increases its runtime by at most a factor of \num{1.5}.
  We provide \jaqsi as open-source software on GitHub and PyPI.
\end{abstract}


\glsresetall

\section{Introduction}

Quantum hardware remains a scarce resource as access to devices is limited and, in the \gls{nisq} era, their results are dominated by noise~\cite{preskill_quantum_2018}.
Classical simulation is therefore still the primary workbench on which quantum algorithms are developed and validated.
It also allows inspection of the full quantum state, controlled use of noise and gradients through the simulated state.
This is especially relevant in \gls{qml}, where \glspl{pqc} are trained by classical optimisers~\cite{mitarai_quantum_2018,benedetti_parameterized_2019,cerezo_challenges_2022}.
A single training run may evaluate the circuit thousands of times, often requiring a gradient at each step.
What usually requires the parameter-shift rules~\cite{wierichs_general_2022} can instead be done by backpropagation through the state in simulation.
However, exact statevector simulation grows exponentially with the number of qubits~\cite{zhou_what_2020}, so simulation throughput constrains the size of the systems examined in studies.

General-purpose frameworks such as Qiskit~\cite{javadiabhari_quantum_2024}, Qibo~\cite{efthymiou_qibo_2022} and Qulacs~\cite{suzuki_qulacs_2021} focus on fast circuit execution, and PennyLane~\cite{bergholm_pennylane_2018} established the circuit as a differentiable node in a hybrid computation.
At the other end, quantum dynamics libraries such as QuTiP~\cite{li_pulse_2022,lambert_qutip_2026} integrate the Schr\"odinger and master equations under time-dependent Hamiltonians, which is the level at which control pulses act on the physical qubits.
Building such tools on JAX~\cite{bradbury_jax_2018} is not new either.
At gate level, TensorCircuit~\cite{zhang_tensorcircuit_2023} offers differentiable and vectorised simulation through tensor-network contraction.
At pulse level, Qiskit Dynamics~\cite{puzzuoli_qiskit_2023}, dynamiqs~\cite{guilmin_dynamiqs_2025} and JAXQuantum~\cite{jha_jaxquantum_2024} provide differentiable, hardware-accelerated solvers for quantum dynamics, and the pulse module of PennyLane relies on JAX as well.
However, these tools leave room for a single stack that covers both gate and pulse levels while supporting end-to-end differentiation, vectorised batches, cached compilation and a plain Python interface.
\autoref{tab:simulators} summarises these features for the simulators compared in \autoref{sec:benchmarking}.

\begin{table}[htb]
  \centering
  \caption{Feature coverage of the simulators considered in this work.
    Gate and Pulse denote simulation levels.
    Grad, Batch and GPU denote gradient computation, vectorised batch execution and GPU acceleration, respectively.
    (\cmark) marks support limited to a backend, add-on or execution mode, such as Qibo's pulse hardware emulator, QuTiP's JAX-based gradients or PennyLane's gate-level parameter broadcasting.}
  \label{tab:simulators}
  \begin{tblr}{colspec={lccccc}, row{1}={font=\bfseries}}
    \toprule
    Simulator & Gate & Pulse & Grad & Batch & GPU \\
    \midrule
    PennyLane~\cite{bergholm_pennylane_2018}  & \cmark   & \cmark   & \cmark   & (\cmark) & \cmark   \\
    Qiskit~\cite{javadiabhari_quantum_2024}   & \cmark   & \xmark   & \xmark   & (\cmark) & \cmark   \\
    Qibo~\cite{efthymiou_qibo_2022}           & \cmark   & (\cmark) & (\cmark) & \xmark   & \cmark   \\
    Qulacs~\cite{suzuki_qulacs_2021}          & \cmark   & \xmark   & (\cmark) & \xmark   & \cmark   \\
    QuTiP~\cite{lambert_qutip_2026}           & (\cmark) & \cmark   & (\cmark) & \xmark   & (\cmark) \\
    dynamiqs~\cite{guilmin_dynamiqs_2025}     & \xmark   & \cmark   & \cmark   & \cmark   & \cmark   \\
    \jaqsi                                    & \cmark   & \cmark   & \cmark   & \cmark   & \cmark \\
    \bottomrule
  \end{tblr}
\end{table}

Combining pulse-synthesised gates, ideal gates and noise channels in one circuit allows the pulse parameters of one layer to be trained while the other gates remain ideal and noise acts only on selected qubits.
This isolates the effect of pulse-level control within one differentiable model.

The contributions can be summarised as follows:
\begin{enumerate*}[label=(\arabic*)]
  \item a unified gate- and pulse-level simulator in which pulse-synthesised gates, unitary gates and noise channels compose freely within one circuit,
  \item an execution engine with cached, batch-vectorised executables that reduces state passes through fusion and gate structure, keeps statevector storage independent of circuit depth during adjoint differentiation, and reduces pulse solves through closed-form, deduplicated and vectorised integration, and
  \item a benchmark comparing \jaqsi against PennyLane, Qiskit, Qibo, Qulacs, QuTiP and dynamiqs at both gate and pulse levels.
\end{enumerate*}

\jaqsi is the stand-alone continuation of the simulation engine behind \texttt{qml-essentials}~\cite{strobl_qml_2025}, which was developed to evaluate thousands of parameter samples in studies of \glspl{qfm}~\cite{perezsalinas_data_2020,schuld_effect_2021,strobl_fourier_2026,strobl_unflattening_2026}.
Up to version 0.2.0~\cite{franz_qml_2026}, the engine was part of \texttt{qml-essentials}.
From version 0.3.0~\cite{franz_qml_2026_2} on, \texttt{qml-essentials} uses \jaqsi as a separate package.
Its pulse-level modelling and \gls{qoc} were utilised in Ref.~\cite{franz_software_2026} and carried the experiments of Ref.~\cite{strobl_beyond_2026}, which show that access to pulse parameters improves the training of \glspl{qfm}.
We provide \jaqsi under an MIT licence on GitHub\footnote{\geneurl{https://github.com/cirKITers/jaqsi}} and through PyPI.

\section{Architecture}
\label{sec:architecture}

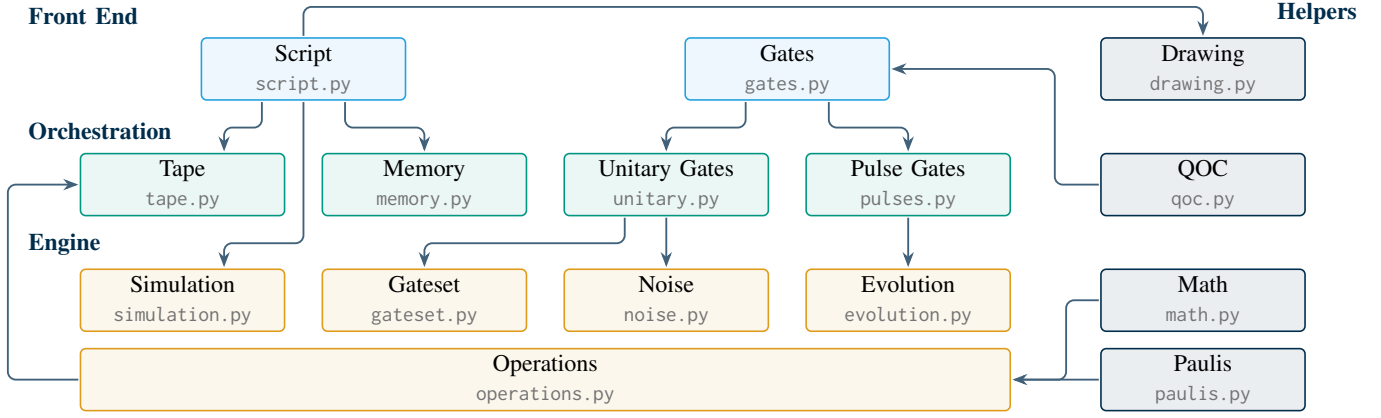
\begin{figure*}[t]
  \centering
  \input{figures/jaqsi_overview.tex}
  \caption{Architectural overview of \jaqsi.
    The front end records circuits on a tape and routes gates to their unitary or pulse implementations.
    The engine applies the resulting operations, which share an operation layer.
    Helper modules provide drawing, \gls{qoc} and analysis.
    Arrows denote the principal dependencies between modules.}
  \label{fig:jaqsi_arch}
\end{figure*}

\jaqsi is organised in four layers:
\begin{enumerate*}[label=(\roman*)]
  \item a \emph{front end} that records quantum circuits and dispatches gates,
  \item an \emph{orchestration} layer holding the recording tape, the memory model for batch chunking and the unitary and pulse implementations of every gate,
  \item an \emph{engine} layer with the simulation kernels, the gate library, the noise channels and the Hamiltonian time evolution, all built on a shared \texttt{Operation} base, and
  \item \emph{helper} modules for drawing, \gls{qoc} and analysis.
\end{enumerate*}
An overview of the architecture is shown in~\autoref{fig:jaqsi_arch}.
The entire stack is implemented in JAX~\cite{bradbury_jax_2018}, making every simulation a pure function of its numerical inputs and fully differentiable.

\subsection{Usage}
\label{sec:arch_usage}

From the user's perspective, a circuit is a plain Python function that applies gates through the \texttt{Gates} entry point, and \texttt{Script} wraps it for execution.
Wires are declared per gate, either as a single index or as a list for multi-qubit gates:

\begin{lstlisting}[language=Python]
  import jaqsi
  from jaqsi import Gates
  import jax.numpy as jnp

  def circuit(phi, theta, omega):
    Gates.Rot(phi, theta, omega, wires=0)
    Gates.CX(wires=[0, 1])
    Gates.CX(wires=[1, 2])

  script = jaqsi.Script(circuit, n_qubits=3)
  res = script.execute(type="expval", obs=[jaqsi.PauliZ(wires=0), jaqsi.PauliZ(wires=2)], args=(jnp.pi, jnp.pi/3, jnp.pi/4))
\end{lstlisting}

Arguments reach the circuit function through \texttt{execute}, while \texttt{type} and the observables in \texttt{obs} select the measurement.
The following subsections describe what happens behind this call, from recording the circuit to executing it at gate and pulse level.

\subsection{Circuit Recording}
\label{sec:arch_frontend}

A quantum circuit in \jaqsi is a plain Python function whose body instantiates an \texttt{Operation} to be registered via a \emph{tape} mechanism.
The tape is a thread-local stack of recorded operations that supports nested recordings and is rebuilt on every trace.
The interface for the user can be divided into two classes:
\begin{enumerate*}[label=(\roman*)]
  \item \texttt{Gates} is the entry point for applying gates: it dispatches each call either to its unitary implementation (\texttt{UnitaryGates}), or, in pulse mode, to its pulse implementation (\texttt{PulseGates}), and attaches noise channels when configured to do so.
  \item \texttt{Script} wraps the circuit function and is the entry point for everything that consumes a recorded tape, most importantly the execution described in \autoref{sec:arch_engine}.
\end{enumerate*}

\subsection{Operation Layer}
\label{sec:arch_operations}

Gates, observables and noise channels derive from the \texttt{Operation} base class.
This class stores target wires and parameters, supports symbolic composition (\texttt{dagger}, \texttt{power}, products) and provides the tensor-contraction machinery.
The specific gate implementation resides in the \texttt{gateset} module and contains, besides the usual fixed and parameterised gates, a virtual \texttt{Barrier} and a diagonal qubit unitary.
Noise is expressed through Kraus channels~\cite{nielsen_quantum_2010} covering bit and phase flip, depolarising, amplitude and phase damping, thermal relaxation and generic user-defined channels, implemented in the \texttt{Noise} module.
The operation layer also defines the Hamiltonian types feeding the pulse engine with \texttt{ParametrizedHamiltonian} representing a sum of time-dependent coefficient functions times \texttt{Hermitian} terms.
A \texttt{Hermitian} wraps a static generator or observable.
Both expose an \texttt{evolve} method returning a gate factory for the pulse-level engine of \autoref{sec:arch_pulse}.

A design decision central to performance is that an operation only ever stores the $2^k \times 2^k$ matrix acting on its $k$ target qubits.
Full $2^n \times 2^n$ matrices are never materialised for gates and channels, and observables are lifted to the full Hilbert space only for measurements outside the fast path for diagonal single-qubit observables of \autoref{sec:arch_engine}.
In addition, each gate declares static properties of its matrix, namely whether it permutes the computational basis (\eg \X, \CX, SWAP), is diagonal (\eg \Z, $\RZ$, \CZ), acts only if its leading control qubits are set (\eg \CRX), and is unitary.
This information is then used in the execution engine instead of inspecting the matrix entries, which are usually traced and thus unknown when the circuit is compiled.

\subsection{Execution Engine}
\label{sec:arch_engine}

A call to \texttt{Script.execute} proceeds in four stages:
\begin{enumerate*}[label=(\arabic*)]
  \item The circuit function is run once so that each operation registers itself on a fresh tape.
  \item The qubit count is inferred from the recorded wires, and the presence of a Kraus channel on the tape selects density-matrix over statevector simulation.
  \item The recorded operations are compiled into an execution plan, which one of the stateless simulation kernels applies to the state without any Python dispatch per gate.
  Whenever a density matrix is requested for a noise-free circuit, the engine simulates the statevector instead and forms $\rho = \ket{\psi}\!\bra{\psi}$ once, reducing the per-gate cost from $O(4^n)$ to $O(2^n)$.
  \item The resulting state is reduced to the requested output (\texttt{state}, \texttt{probs}, \texttt{expval} or \texttt{density}) and optionally sampled from if shots are required.
  Diagonal single-qubit observables take a fast path that reduces the probability vector in $O(2^n)$ per observable instead of contracting a lifted $2^n \times 2^n$ observable at $O(4^n)$.
\end{enumerate*}

\jaqsi natively supports batched execution over parameters, inputs, or initial states.
Here, any argument axis can be declared as a batch dimension, in which case the engine vectorises the simulation kernel via \texttt{vmap} and caches the resulting execution plan.
The batched execution function is wrapped with Equinox's filtered \gls{jit}~\cite{kidger_equinox_2021}, which treats Python values as static and reuses compiled code for matching signatures.
Purely array-valued signatures are additionally compiled ahead of time into an \gls{xla} executable which is then invoked directly.
This allows skipping the per-call dispatch overhead usually dominating small circuits in tight loops.
Before launching a batch, a memory model estimates the peak requirement from a fixed number of state-sized buffers, independent of circuit depth, and splits the batch into chunks if it exceeds the given budget, which requires that a single circuit alone fits into memory.
On CPU, a batch whose working set exceeds the share of the last-level cache per core is further processed in vectorised tiles that fit into it.
On CPU, \texttt{JAX\_NUM\_CPU\_DEVICES} sets the number of JAX devices.
Sufficiently large batches are sharded when multiple devices are available and the batch size is divisible by their number.

On top of this compiled and vectorised execution, the engine lowers the cost of a circuit in three ways.
It reduces passes over the state, bounds gradient memory independently of circuit depth and, at pulse level, reduces the number of solves.
The first lever acts in the statevector kernel, which keeps the state in rank-$n$ tensor form, so that every entry of the plan costs one pass over the $2^n$ amplitudes.
To reduce the number of these passes, consecutive gates on the same wires are multiplied into a single matrix as long as no gate in between touches these wires, similar to the gate merging of Qulacs~\cite{suzuki_qulacs_2021} and the gate clustering of Ref.~\cite{haner_0_2017}.
As these small matrix products remain part of the traced program, gradients flow through them unchanged.

The static gate properties of \autoref{sec:arch_operations} determine how each plan entry updates the state.
Permutations reindex the amplitudes, diagonal gates scale them elementwise, and controlled gates act only on amplitudes for which their controls are set.
Runs of consecutive permutations or diagonal gates on arbitrary wires are also composed into a single precomputed index gather or a single elementwise factor, respectively.
For the hardware-efficient ansatz of \autoref{fig:benchmark_circuit_gate} on $n$ qubits, this turns the $4n$ gates into $n$ dense single-qubit blocks and one gather for the entire \CX{} ring.
Remaining dense blocks on one or two qubits are applied as weighted sums of state slices and wider ones through \texttt{einsum}.

The same lever applies to the density-matrix kernel, which treats $\rho$ as a rank-$2n$ tensor with separate ket and bra axes.
A gate on up to two qubits enters the plan as its superoperator $U \otimes U^*$ and a channel as $\sum_k K_k \otimes K_k^*$~\cite{wood_tensor_2015}, both acting on the ket and bra axes of their wires at once.
Gates and channels thereby become the same kind of plan entry and are fused alike, so that each block costs one pass over $\rho$ instead of two per gate and two per Kraus operator.
In the noisy variant of the benchmark circuit (\autoref{sec:bench_setup}), for instance, the $9n$ gates and channels collapse into $4n$ blocks.

The second lever bounds the memory of gradients.
For exact expectation values of circuits whose gates are all unitary, the engine computes gradients with the adjoint method~\cite{jones_efficient_2020}, implemented as a custom vector-Jacobian product in JAX.
Its forward pass keeps only the final state and the matrices of the plan, and its backward pass walks the plan in reverse, uncomputing the state with $U^\dagger$ while propagating the cotangent.
This bounds the memory to a constant number of statevectors regardless of circuit depth, similar to the reversible differentiation in Yao~\cite{luo_yao_2020}.
As the cotangents refer to the fused matrices, JAX chains them into the gate parameters without requiring a generator per gate.
All other cases fall back to automatic differentiation through the kernel, which stores the intermediate states.
Either way, the simulation remains a pure JAX function of its arguments, so that gradient-based optimisation with standard optimisers applies directly to the circuit of \autoref{sec:arch_usage}:

\begin{lstlisting}[language=Python]
  import jax
  import optax as otx

  def cost(params):
    return script.execute(type="expval", obs=[jaqsi.PauliZ(wires=0)], args=tuple(params))[0]

  params = jnp.array([0.1, 0.2, 0.3])
  opt = otx.adam(0.1)
  opt_state = opt.init(params)
  for step in range(100):
    grads = jax.grad(cost)(params)
    updates, opt_state = opt.update(grads, opt_state, params)
    params = otx.apply_updates(params, updates)
\end{lstlisting}

\subsection{Pulse-Level Engine}
\label{sec:arch_pulse}

At pulse level, a gate's drive Hamiltonian determines how its propagator changes during the pulse.
\jaqsi obtains that propagator by integrating the Schr\"odinger equation $\mathrm{d}U/\mathrm{d}t = -\imag \hamiltonian(t)\, U$ for the gate's parameterised Hamiltonian.
\texttt{PulseGates} holds the pulse-level composition where leaf gates ($\RX, \RY, \RZ, \CZ$)~\cite{franz_software_2026} carry their own drive Hamiltonian and envelope, while composite gates decompose into sequences of leaves.
Pulse shapes are described by \texttt{PulseEnvelope} (Gaussian, square, cosine, \gls{drag}, sech or user-defined), per-gate parameters by \texttt{PulseParams}, and both are unified in \texttt{PulseInformation} for fine-grained control.
Each envelope $E(t)$ is centred at the midpoint of its pulse and, except for \gls{drag}, modulates a single quadrature of the carrier, whose phase selects the rotation axis of $\RX$ or $\RY$.
As in Qiskit\footnote{\url{https://quantum.cloud.ibm.com/docs/api/qiskit/1.4/qiskit.pulse.library.Gaussian}}, the Gaussian is lifted by its value at the pulse boundaries and rescaled to its peak, so that it vanishes at both ends instead of switching on and off with a step.
Following Ref.~\cite{motzoi_simple_2009}, \gls{drag} adds the derivative $-\beta \dot{E}(t)$ of this Gaussian as a second control on the orthogonal quadrature, where the weight $\beta$ takes the place of the inverse anharmonicity~\cite{gambetta_analytic_2011}.
As \jaqsi models qubits as two-level systems without a leakage level to suppress, and the Gaussian alone already yields the exact rotation under the \gls{rwa}, the calibrated weight vanishes.

The integration itself lives in the \texttt{Evolution} class, which turns a Hamiltonian into a gate factory whose product is a regular \texttt{Operation}.
This makes the execution engine of \autoref{sec:arch_engine} agnostic to whether a gate was defined by a matrix or by a pulse.
Here, two solver families are available in \jaqsi for pulse simulation:
\begin{enumerate*}[label=(\roman*)]
  \item adaptive eighth-order \gls{dp} integrators via diffrax~\cite{kidger_neural_2021}, whose step-size control targets the requested error tolerances for arbitrary envelopes, and
  \item fixed-step commutator-free Magnus integrators of second and fourth order~\cite{blanes_fourth_2006}, which preserve unitarity by construction and incur a static, JIT-friendly cost per gate.
\end{enumerate*}
The third lever reduces the cost of integrating pulses whose Hamiltonian has a single time-dependent drive term.
A key observation here is that for such a pulse, only the integral of the drive coefficient, or pulse area, needs to be evaluated while the matrix part can be reused.
Specifically, if the Hamiltonian is $f(\boldsymbol{p}, t)\,\hamiltonian$ with a constant matrix $\hamiltonian = V \Lambda V^\dagger$, it commutes with itself at all times.
The first term of the Magnus expansion is then exact~\cite{blanes_magnus_2009}: $U = V \exp(-\imag F \Lambda)\, V^\dagger$, where $F = \int f(\boldsymbol{p}, t)\,\mathrm{d}t$ describes the pulse area.
In this case, the adaptive solvers integrate only the scalar $F$, while the eigendecomposition of $\hamiltonian$ is computed once when the circuit is traced.
This keeps $U$ differentiable in the pulse parameters even at degenerate eigenvalues.
Under the \gls{rwa}, the drives of $\RX$ and $\RY$ reduce to such a term for single-quadrature envelopes like the Gaussian, so that every leaf gate takes this path.
A \gls{drag} pulse, in contrast, changes the ratio of its two quadratures over time, does not commute with itself at different times and is therefore integrated as the matrix equation.
Before state evolution, pending pulses with compatible solver settings are grouped into vectorised solves.
Identical pulses within a group, such as the fixed rotations within a Hadamard, are evaluated once, and compiled solvers are reused across gates of the same structure.
Their propagators then enter the ordinary gate-fusion pass of the execution engine of \autoref{sec:arch_engine}.
The shipped default pulse parameters are pre-calibrated against the ideal unitaries by the included \gls{qoc} module, which builds on optax~\cite{deepmind_deepmind_2020} and was presented, together with the drawing capabilities, in Ref.~\cite{franz_software_2026}.

Because \texttt{PulseGates} builds directly on \jaqsi operations, pulse-level gates can be placed in the same circuit as gate-level operations and noise channels, which route the circuit to the density-matrix kernel of \autoref{sec:arch_engine}.
The underlying Hamiltonian time evolution is exposed as well, such that any \texttt{Hermitian} provides an \texttt{evolve} method returning a time-evolution gate factory $U(t) = \exp(-\imag t \hamiltonian)$.
Multiplying a coefficient callable $f(\boldsymbol{p}, t)$ with a \texttt{Hermitian} then yields a \texttt{ParametrizedHamiltonian} with the latter being compiled once through \texttt{evolve}.
Its parameters enter at call time, allowing the resulting operation to serve as a trainable layer in a pulse-level model.
All three mechanisms compose freely in a single circuit:

\begin{lstlisting}[language=Python]
  from jaqsi import PulseGates

  def coeff(p, t):
    return p

  def circuit(w, p):
    PulseGates.RX(w, wires=0)
    PulseGates.CX(wires=[0, 1])
    X = jaqsi.PauliX(record=False).matrix
    ph = coeff * jaqsi.Hermitian(matrix=X, wires=1, record=False)
    ph.evolve()([p], 1.0)
    jaqsi.DepolarizingChannel(0.1, wires=1)

  script = jaqsi.Script(circuit, n_qubits=2)
  rho = script.execute(type="density", args=(jnp.pi * 0.5, 0.3))
\end{lstlisting}

\subsection{Analysis Tools}
\label{sec:arch_tools}

Three auxiliary modules complete the package.
\begin{enumerate*}[label=(\roman*)]
  \item \texttt{math} provides quantum-information utilities such as fidelity, trace distance, the Fubini--Study metric and quantum Fisher information, and partial traces.
  \item \texttt{paulis} implements a symplectic tableau representation of Pauli words~\cite{aaronson_improved_2004} with symbolic Clifford conjugation in $O(n)$ per gate and statevector expectations in $O(n 2^n)$, backing analytical, non-simulative evaluation of circuit classes built on top of \jaqsi~\cite{strobl_qml_2025}.
  \item \texttt{drawing} renders a recorded circuit as ASCII text, a matplotlib figure or \texttt{quantikz} \LaTeX{} code~\cite{franz_software_2026}, and its \texttt{pulse} mode plots the pulse schedule, \ie the sequence of pulses applied on each qubit, where shaded areas denote the pulse envelope and composite gates are automatically decomposed into basis gates.
\end{enumerate*}

Note that while these modules are not required for circuit execution, they provide convenient functionality typically seen in related simulators~\cite{bergholm_pennylane_2018,efthymiou_qibo_2022}.

\section{Benchmarking}
\label{sec:benchmarking}

\subsection{Benchmark Design}
\label{sec:bench_setup}

The benchmark measures the wall-clock time of a batched execution or gradient evaluation against a set of established simulators.
At gate level, it uses the hardware-efficient ansatz of \autoref{fig:benchmark_circuit_gate}.
This circuit follows the Yao benchmark suite~\cite{luo_yao_2020}, and its input encoding layer mirrors the parameter layout of a \gls{qml} training step.
At pulse level, it uses the \hadamard{} and \CRX$(\phi)$ ring circuit of \autoref{fig:benchmark_circuit_pulse}, which the transcribed pulse model covers.
All angles are drawn uniformly from $[-\pi, \pi]$ with a fixed seed and shared across all simulators.
Inputs differ for each instance in a batch, weights are shared across it, and both are redrawn in each iteration.

Gate-level runs span \numrange{2}{12} qubits in every measured mode.
The comparison includes PennyLane~\cite{bergholm_pennylane_2018}, Qiskit~\cite{javadiabhari_quantum_2024}, Qibo~\cite{efthymiou_qibo_2022} and Qulacs~\cite{suzuki_qulacs_2021}.
Each gate-level timing averages \num{20} iterations with a batch size of \num{10}.
At pulse level, the comparison includes PennyLane, QuTiP~\cite{lambert_qutip_2026} and dynamiqs~\cite{guilmin_dynamiqs_2025} for \numrange{2}{12} qubits in each measured mode, over \num{20} iterations with a batch size of \num{2}.
Qiskit dropped its pulse support in version 2.0, Qulacs exposes none, and Qibo does so only through its hardware emulator.

The measured modes at both levels are statevector expectation values (\texttt{expval}), noise-free density matrices (\texttt{density}) and noisy density matrices (\texttt{noise}).
In \texttt{noise}, a depolarising channel of strength $p = \num{0.01}$ follows each gate on every qubit it acts on: $\rho \mapsto (1-p)\rho + \frac{p}{3}(\X\rho\X + \Y\rho\Y + \Z\rho\Z)$.
These channels are part of the circuit, so all simulators apply the same sequence independently of their framework-specific noise models.
Unlike in \texttt{density}, none can derive the noisy result from the statevector.
At pulse level, the channels act between the pulse sequences of consecutive gates.
At gate level, \texttt{grad} measures the gradient of $\sum_i \langle \Z_i \rangle$, summed over the batch, with respect to the weights for all simulators except Qiskit and Qibo.
The results of each simulator are cross-validated against \jaqsi to an absolute tolerance of \num{e-8} at gate level, \num{e-6} for gradients, whose summation order differs between differentiation methods, and \num{e-5} at pulse level, the latter accounting for the accumulated solver error.

The benchmarks ran on a bare-metal node with AVX2 and AVX-512, running NixOS 26.05 (Yarara) at kernel 7.2.6, with \num{8} Intel Xeon Platinum 8260L CPUs (\num{192} cores, \num{384} hardware threads) and \num{5.9} TiB memory.
They ran inside a Docker container based on Debian 12 (bookworm) with Python 3.12, pinned to \num{240} logical CPUs (\num{120} cores with SMT across \num{5} sockets) and limited to \num{4} TiB memory without swap.
Every simulator is pinned to a single vCPU via the process CPU affinity, and multi-threading in OpenMP, BLAS, Qulacs and \gls{xla} is disabled through the respective environment variables and flags.
The software stack uses Python 3.12.14, \jaqsi 0.1.1~\cite{franz_jaqsi_2026}, JAX 0.11.1 and diffrax 0.7.2.
The competitor versions are PennyLane 0.45.1 with Lightning 0.45.0, Qiskit 2.5.2 with Aer 0.17.2, Qibo 0.3.4 with qibojit 0.1.16, Qulacs 0.6.14, QuTiP 5.3.1 and dynamiqs 0.3.6 at commit \texttt{991fd45}.\footnote{The code, configurations and raw results to reproduce the benchmarks are available at \geneurl{https://github.com/cirKITers/jaqsi-data}~\cite{strobl_jaqsi_2026}.}

Circuit and device construction happen outside the timed loop with garbage collection being disabled inside the loop.
A single untimed warmup pass triggers \gls{jit} compilation.
The reported values are the mean and standard deviation of the time per batch over the timed iterations.
All simulators run in their performance-optimised configuration, explained in detail below.
PennyLane executes a jitted QNode on \texttt{default.qubit}, or on \texttt{default.mixed} in the \texttt{noise} mode, which vectorises the batch through parameter broadcasting like \jaqsi and is differentiated by backpropagation through JAX.
As a second gradient reference, PennyLane Lightning runs the adjoint method~\cite{jones_efficient_2020} on the compiled \texttt{lightning.qubit} device, which executes a broadcast batch sample by sample and accepts no channels.
Qiskit runs on Aer with the \texttt{statevector} method in \texttt{expval} and the \texttt{density\_matrix} method otherwise, submitting the batch as a single parameterised circuit whose parameters Aer binds per sample\footnote{\url{https://qiskit.github.io/qiskit-aer/stubs/qiskit_aer.AerSimulator.html}}.
Qibo runs on its qibojit backend in the measured modes and evaluates each $\langle \Z_i \rangle$ through its symbolic observable\footnote{\url{https://qibo.science/qibo/stable/code-examples/advancedexamples.html}}.
As gate fusion requires concrete angles, Qulacs rebuilds and fuses its circuit for every sample inside the timed loop.
In these measurements, this is faster than updating the angles of a parametric circuit built once from \num{10} qubits on in \texttt{expval} and from \num{4} in \texttt{density}.
In \texttt{noise}, where its optimiser would discard the channels, Qulacs runs unfused and builds its circuit only once.
Qibo and Qulacs loop over the batch in Python, and Qulacs differentiates each sample through the backpropagation of its parametric circuits, whereas Qiskit and Qibo require additional packages for gradients.
Qiskit and Qibo parameterise the depolarising channel by the weight $\lambda$ of the maximally mixed state, so the benchmark sets $\lambda = 4p/3$~\cite{nielsen_quantum_2010}.

At pulse level, each gate expands into a sequence of time evolutions $\mathrm{d}U/\mathrm{d}t = -\imag \hamiltonian(t) U$, one per basis gate of its decomposition.
A \hadamard{} contributes \num{3} such pulses and a \CX{} \num{7}, so the \CRX{} of \autoref{fig:benchmark_circuit_crx} accounts for \num{18} and the full circuit for $21n$.
That schedule is transcribed once from \jaqsi, using its calibrated Gaussian envelope for driven rotations under the \gls{rwa}, so that all backends integrate identical pulses instead of their own pulse model.
Each pulse is integrated on its own two- or four-dimensional space and the resulting propagator is contracted into the statevector, or applied from both sides to the density matrix in the \texttt{noise} mode, as in the pulse-level engine of \autoref{sec:arch_pulse}.
In \texttt{density}, all backends form the density matrix from the final statevector.
In \texttt{noise}, PennyLane applies channels through \texttt{default.mixed}, while the benchmark code applies them for QuTiP and dynamiqs.
\jaqsi and dynamiqs vectorise the batch dimension while PennyLane and QuTiP loop over it in Python.
Furthermore, PennyLane integrates each pulse directly with the \gls{dp} 5(4) solver of \texttt{jax.experimental.ode} that its \texttt{ParametrizedEvolution} is built on, bypassing the per-operation overhead of that API, QuTiP with its default Adams method, and \jaqsi and dynamiqs with \gls{dp} 8(7).
As every pulse Hamiltonian is a single term $f(t)\,\hamiltonian$ that commutes with itself at all times, \jaqsi integrates only the scalar pulse area $\int f(t)\,\mathrm{d}t$ and exponentiates $\hamiltonian$ in its eigenbasis~\cite{blanes_magnus_2009}, whereas the other backends integrate the full matrix equation.
All backends run at an absolute and relative solver tolerance of \num{e-10}.

The transcription is validated by comparing the infidelity between pulse- and gate-level simulation for each simulator, with \jaqsi as the gate-level reference for QuTiP and dynamiqs.
Across both pulse configurations, all backends and all qubit counts, the infidelities of the noise-free and noisy density matrices range from \numrange{7e-15}{6e-14}, within a few hundred times double-precision machine epsilon.
All four backends reach nearly the same value at each qubit count, consistent with a residual from the shared pulse schedule.
Gate- and pulse-level simulation therefore agree to numerical precision.

\begin{figure}[t]
  \centering
  \input{figures/benchmark_circuit.tex}
  \caption{Benchmark circuits for $n = 3$.
    (a) Gate-level hardware-efficient ansatz with input angles $x$ and trainable weights $w$.
    (b) Pulse-level circuit with a \CRX$(\phi)$ ring that closes from the last qubit to the first.
    (c) Decomposition of one \CRX{} into rotations and two \CX{} gates, with control $c$ and target $t$.
    Each \CX{} becomes $\hadamard \CZ \hadamard$ on its target, and each \hadamard{} becomes $\RZ(\pi)$, $\RY(\pi/2)$ and a correction phase.
    The resulting circuit has $21n$ pulses.}
  \label{fig:benchmark_circuit}
\end{figure}
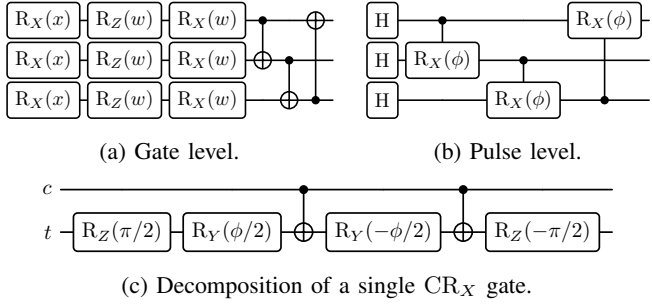

\subsection{Results}
\label{sec:bench_results}

\begin{figure}[htb]
  \centering
  \resizebox{\columnwidth}{!}{\input{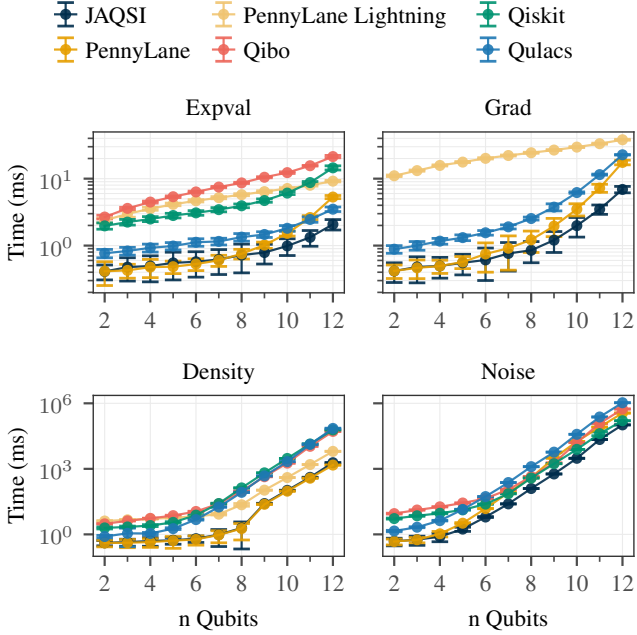}}
  \caption{Gate-level execution time versus qubit count for the single-layer hardware-efficient ansatz, with a logarithmic time axis.
    Expval denotes statevector expectation values, Grad denotes gradients, and Density and Noise denote noise-free and noisy density matrices.
    Each point is the mean time per batch of \num{10} over \num{20} iterations.
    Error bars denote the standard deviation.
    Qiskit and Qibo are absent from Grad, and PennyLane Lightning is absent from Noise.}
  \label{fig:benchmarks-absolute-gate}
\end{figure}

\autoref{fig:benchmarks-absolute-gate} shows two runtime regimes at gate level: fixed execution overhead dominates small circuits, while passes over the state matter increasingly as the qubit count grows.
\autoref{tab:speedup} summarises the speedup of \jaqsi over the compared simulators at both levels.
Within the timed loop, every call of \jaqsi hits the cached, ahead-of-time compiled \gls{xla} executable and the whole batch is vectorised via \texttt{vmap}.
The jitted PennyLane QNode compiles to an equally batched executable, whereas PennyLane Lightning, Qulacs, Qiskit and Qibo handle the batch sample by sample and pay a fixed cost per sample.
At few qubits, this per-sample overhead dominates, so \jaqsi and PennyLane have similar runtimes and lead the remaining simulators.
As the state grows, passes over it account for more of the runtime, and the optimisations of \autoref{sec:arch_engine} determine the margins.

\begin{table}[htb]
  \centering
  \caption{Speedup of \jaqsi over each simulator across the measured qubit counts.
    Each range gives the minimum and maximum ratio of the other simulator's mean execution time to that of \jaqsi for the indicated mode.
    Values below one mean the other simulator is faster.
    Expval, Density, Noise and Grad denote statevector expectation values, noise-free and noisy density matrices, and gradients, respectively.
    Dashes mark modes without a reported comparison.
    In the pulse-level matrix-ODE rows, \jaqsi integrates the matrix equation instead of using its closed-form pulse solve.}
  \label{tab:speedup}
  \input{figures/speedup.tex}
\end{table}

In statevector simulation, \jaqsi and PennyLane are within about \num{15} percent through \num{7} qubits.
Beyond that, \jaqsi leads by up to a factor of \num{2.6}.
Qulacs trails by \numrange{1.7}{2.0} times, while Qiskit and Qibo trail by \numrange{4.7}{7.0} and \numrange{6.5}{13} times, respectively.
The lead over PennyLane Lightning peaks at \num{7} qubits and narrows beyond.
The growing gap to PennyLane reflects the first lever: fusion reduces the $4n$ gates of the ansatz to $n + 1$ plan entries, whereas PennyLane applies every gate in a pass of its own.

In noise-free density-matrix simulation, \jaqsi is on par with or slightly slower than PennyLane at every measured qubit count, taking at most \num{28} percent longer.
Both form the density matrix from the final statevector, so writing the $4^n$ output dominates from \num{9} qubits on.
PennyLane Lightning trails \jaqsi by \numrange{3.3}{12} times, while Qulacs, Qiskit and Qibo evolve the full density matrix and fall behind by up to \num{68} times.

Under noise, \jaqsi is the fastest from \num{4} qubits on, with a consistent lead of \numrange{1.4}{5.9} times over PennyLane.
Its lead over Qiskit falls from about \num{12} times at \num{4} qubits to \num{1.5} times at \num{12}, while the lead over Qibo remains at least \num{3.2} times across the run.
Every simulator evolves the full density matrix in this mode, so passes over it set the runtime at larger qubit counts.
\jaqsi applies $4n$ fused superoperator blocks, whereas PennyLane's \texttt{default.mixed} applies each of the $9n$ gates and channels separately, which is in line with this lead.
The lead over the remaining simulators at few qubits, in turn, reflects their per-sample overhead.

In gradient evaluation, \jaqsi is on par with PennyLane up to \num{5} qubits and pulls ahead beyond.
Qulacs trails by \numrange{2.1}{3.3} times across the gate-level run.
Qiskit and Qibo require additional packages for gradients and are excluded from this mode.
The lead of \jaqsi over PennyLane Lightning peaks at \num{6} qubits and narrows steadily beyond.
The margin over PennyLane is the second lever: the adjoint method walks the fused plan backwards once, so that a gradient of \jaqsi costs about \num{3} forward passes at \num{12} qubits, whereas backpropagation in PennyLane caches the intermediate state of every gate~\cite{jones_efficient_2020}.
PennyLane Lightning uses the adjoint method as well, but runs the batch sample by sample through the Python interface of PennyLane, so that its per-sample cost dominates up to \num{12} qubits while the time of \jaqsi already almost doubles per qubit.
The extended run up to \num{20} qubits in \autoref{app:scaling} shows how this advantage narrows.
PennyLane Lightning overtakes \jaqsi in statevector simulation from \num{19} qubits on and is faster in gradient evaluation only at \num{19} qubits, while the lead of \jaqsi over PennyLane grows to \num{5.4} times in statevector simulation.

\begin{figure}[htb]
  \centering
  \resizebox{\columnwidth}{!}{\input{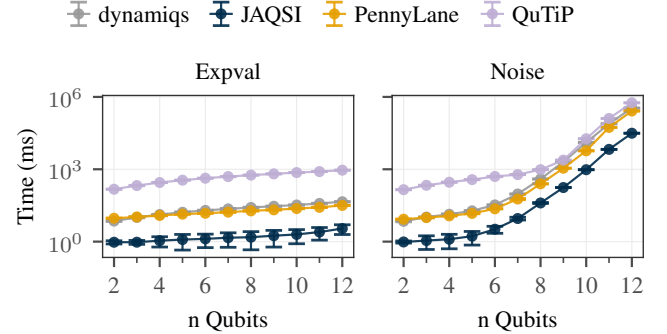}}
  \caption{Pulse-level execution time from \numrange{2}{12} qubits for the \hadamard{} and \CRX{} ring circuit, with a logarithmic time axis.
    Expval denotes statevector expectation values, and Noise denotes noisy density matrices.
    Each execution integrates $21n$ pulse evolutions for $n$ qubits.
    Each point is the mean time per batch of \num{2} over \num{20} iterations.
    Error bars denote the standard deviation.}
  \label{fig:benchmarks-absolute-pulse}
\end{figure}

\autoref{fig:benchmarks-absolute-pulse} shows the pulse-level results.
In statevector simulation, \jaqsi leads PennyLane by a nearly constant factor, dynamiqs by a factor that grows to \num{17} at \num{8} qubits and declines to \num{13} at \num{12}, and QuTiP by more than two orders of magnitude.
Under noise, the leads over PennyLane and dynamiqs persist, whereas the lead over QuTiP drops from above \num{200} times at \num{4} qubits to \numrange{14}{19} times from \num{9} qubits on.
In noise-free density-matrix simulation, the leads resemble those in statevector simulation at few qubits.
From \num{10} qubits on, materialising the $4^n$ density matrix dominates the runtime of \jaqsi.
Its lead then shrinks to \numrange{1.1}{6.3} times at \num{12} qubits.

Part of the gap to PennyLane and QuTiP reflects the batching asymmetry noted in \autoref{sec:bench_setup}, although a batch of \num{2} bounds that share to a factor of \num{2}.
The larger part stems from the third lever, the pulse-level engine of \autoref{sec:arch_pulse}.
As every pulse of the benchmark schedule is a single-term Hamiltonian, \jaqsi integrates the scalar pulse area instead of the matrix equation.
It further stacks all pulses of equal shape into one vectorised solve, solves identical pulses only once, \eg \num{18} distinct solves for the \num{126} pulses at \num{6} qubits, and fuses the resulting propagators, \eg the \num{63} pulses at \num{3} qubits into \num{16} passes over the state.
PennyLane and dynamiqs, in contrast, integrate one matrix equation per pulse in sequence, and QuTiP additionally dispatches every solve through Python.
The lead over PennyLane stays nearly constant with the qubit count, as \jaqsi lowers both the fixed cost per call and the time per additional pulse by about the same factor.
Under noise, the density matrix dominates the runtime from about \num{7} qubits on, where \jaqsi applies pulses and channels as fused superoperators, while the per-solve overhead of QuTiP loses weight.
The matrix-ODE run in \autoref{app:pulse-ode} compares performance with the closed-form solve disabled.
Disabling it increases the runtime of \jaqsi by at most a factor of \num{1.5} in the plotted statevector and noisy density-matrix modes.
\jaqsi remains faster than the compared pulse simulators in these modes (\autoref{tab:speedup}).

\section{Discussion \& Conclusion}
\label{sec:conclusion}

\jaqsi combines gate- and pulse-level simulation in an end-to-end differentiable, batched and compiled stack while keeping circuits as plain Python functions.
In the main single-threaded gate-level benchmark up to \num{12} qubits, it is faster than Qulacs, Qiskit and Qibo in every compared mode at every measured qubit count.
It keeps pace with or outperforms the compiled PennyLane path in statevector simulation and gradient evaluation, and it is the fastest of the compared pulse-level simulators in the plotted statevector and noisy density-matrix modes.
In noise-free density-matrix simulation, \jaqsi is on par with or slightly slower than PennyLane at every measured qubit count.
PennyLane Lightning illustrates how the gate-level margins change with circuit size: its per-sample overhead favours \jaqsi at smaller qubit counts, while the extended run of \autoref{app:scaling} shows it overtaking \jaqsi in statevector simulation from \num{19} qubits on and in gradient evaluation at \num{19} qubits.
Qulacs also closes in on \jaqsi in statevector simulation and is \num{8} percent faster at \num{20} qubits, while remaining slower in gradient evaluation.

Two features of the benchmark design qualify these comparisons.
In the noise-free \texttt{density} mode, \jaqsi and PennyLane form the result from a statevector, whereas Qiskit, Qibo and Qulacs evolve the full $O(4^n)$ density matrix.
The margin over the latter therefore partly reflects the cheaper computation.
In the \texttt{noise} mode, every simulator evolves the full density matrix.
Here the lead of \jaqsi reflects vectorised batches and fused superoperators.

At pulse level, every pulse of the benchmark schedule has a single-term Hamiltonian, which \jaqsi solves through its pulse area while the other backends integrate the matrix equation.
This shortcut only applies to single-quadrature envelopes such as the Gaussian used in the benchmark.
The \gls{drag} envelope of \jaqsi adds its derivative correction as a second control on the orthogonal quadrature~\cite{motzoi_simple_2009}, so that its pulses, as well as drives beyond the \gls{rwa}, do not commute with themselves at different times and fall back to the matrix equation.
Solving that equation for the Gaussian benchmark schedule makes \jaqsi at most \num{1.5} times slower (\autoref{fig:benchmarks-absolute-pulse-ode}).
This comparison suggests that the shortcut accounts for only a small part of its lead in the plotted modes.

The remaining limits concern memory, differentiation and benchmark scope.
Chunking, tiling and sharding act only on the batch axis, so a single state that exceeds the memory budget cannot be executed regardless of batch size.
Representations that exploit limited entanglement, such as tensor networks, could extend this range beyond the statevector approach.
The adjoint gradient uses a custom vector-Jacobian product and therefore supports reverse-mode differentiation only.
Forward-mode differentiation, as used for the quantum Fisher information of the \texttt{math} module, falls back to automatic differentiation through the simulation kernel.
The benchmarks use \num{1} thread, batches of \num{10} at gate level and \num{2} at pulse level, and a single-layer ansatz.
Larger batches, deeper circuits and multi-threaded or GPU execution may shift the margins, so those configurations need separate measurement.
Future work may investigate tensor-network representations, benchmark these other configurations and add an interface through which \jaqsi can serve as a backend for other simulators.

\small
\noindent\textbf{Acknowledgements}
Implementation of numerical experiments as well as grammar and wording checks were partially supported by Anthropic Claude Opus series.
MS, EK and AS acknowledge support by the state of Baden-W\"urttemberg through bwHPC.
LS acknowledges the support by the Doctoral School Karlsruhe School of Elementary and Astroparticle Physics: Science and Technology.
MF and WM acknowledge partial support by the German Federal Ministry of Research, Technology and Space (BMFTR), funding programme \enquote{Research Program Quantum Systems}, grant number 13N17387, and by the German Research Foundation, grant 563122858 (MA 9739/1-1). WM acknowledges support by the High-Tech Agenda of the Free State of Bavaria.
\normalsize

\newpage

\printbibliography
\glsresetall

\newpage

\appendices

\section{Gate-Level Scaling Benchmark}
\label{app:scaling}

To examine gate-level performance beyond \num{12} qubits, the \texttt{expval} and \texttt{grad} modes of \autoref{sec:bench_setup} are repeated for the single-layer hardware-efficient ansatz from \numrange{8}{20} qubits.
Each timing averages \num{50} iterations with a batch size of \num{10}.
This run includes PennyLane, PennyLane Lightning and Qulacs alongside \jaqsi.
While the runtimes of \jaqsi and PennyLane already grow by about a factor of \num{2} per qubit at \num{12} qubits, that of PennyLane Lightning only does so once the state outweighs its fixed cost per sample, so that a crossover is expected at larger qubit counts.

\begin{figure}[htb]
  \centering
  \resizebox{\columnwidth}{!}{\input{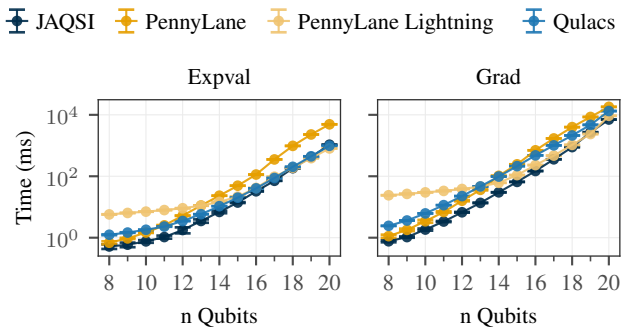}}
  \caption{Gate-level execution time of \jaqsi, PennyLane, PennyLane Lightning and Qulacs from \numrange{8}{20} qubits for statevector expectation values (Expval) and gradients (Grad) of the single-layer hardware-efficient ansatz, with a logarithmic time axis.
    Each point is the mean time per batch of \num{10} over \num{50} iterations.
    Error bars denote the standard deviation.}
  \label{fig:benchmarks-absolute-scaling}
\end{figure}

\autoref{fig:benchmarks-absolute-scaling} confirms this expectation.
In statevector simulation, the lead of \jaqsi over PennyLane Lightning narrows from \num{11} times at \num{8} qubits to \num{3} percent at \num{18} qubits.
PennyLane Lightning is \num{1.1} times faster at \num{19} qubits and \num{1.3} times faster at \num{20} qubits.
Meanwhile, the lead over PennyLane grows from \num{1.3} times at \num{8} qubits to \numrange{4.6}{5.4} times from \num{17} qubits on.
Qulacs closes its gap to \jaqsi from \num{2.4} times at \num{8} qubits to \num{6} percent at \num{19} qubits and is \num{8} percent faster at \num{20} qubits.
From \num{18} qubits on, the time of \jaqsi grows by \numrange{2.3}{2.6} times per qubit, whereas that of PennyLane Lightning about doubles, so PennyLane Lightning is expected to extend its lead beyond \num{20} qubits in this mode.
In gradient evaluation, the lead of \jaqsi over PennyLane Lightning narrows from \num{31} times at \num{8} qubits to \num{1.2} times at \num{18} qubits.
PennyLane Lightning is \num{1.1} times faster at \num{19} qubits, but its runtime grows about fourfold from \numrange{19}{20} qubits, so that \jaqsi leads again by \num{1.3} times at \num{20} qubits.
\jaqsi remains \numrange{1.5}{4.8} times faster than PennyLane.
It also remains ahead of Qulacs in gradients, although that lead narrows from \num{3.2} times at \num{8} qubits to \numrange{1.8}{1.9} times at \num{19} and \num{20} qubits.
Here, the cost of a gradient of \jaqsi keeps rising with the qubit count and reaches about \num{7} forward passes at \num{20} qubits.

\newpage

\section{Pulse-Level Matrix-ODE Benchmark}
\label{app:pulse-ode}

To assess the role of its closed-form pulse solve, \jaqsi is run again on the pulse schedule of \autoref{sec:bench_setup} with that solve disabled.
The run spans \numrange{2}{12} qubits with batches of \num{2} over \num{20} iterations.
It integrates the full matrix equation for each pulse, as PennyLane, QuTiP and dynamiqs do in both runs.
The other backends and the absolute and relative solver tolerances of \num{e-10} remain the same.
Both configurations use the default Gaussian envelope of \jaqsi for driven rotations.
The second quadrature of \gls{drag} would require the matrix equation for every $\RY$ in either configuration, so the Gaussian keeps the pulse schedule otherwise unchanged.

\begin{figure}[htb]
  \centering
  \resizebox{\columnwidth}{!}{\input{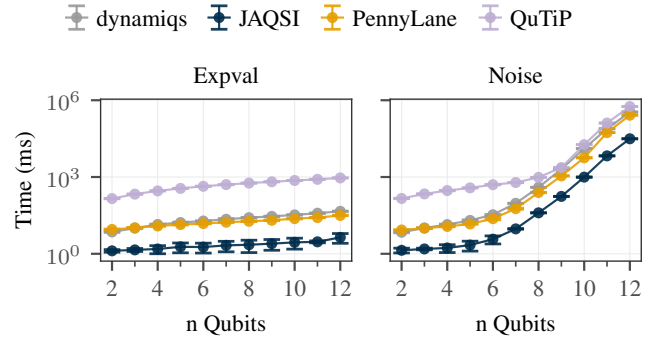}}
  \caption{Pulse-level execution time from \numrange{2}{12} qubits for the \hadamard{} and \CRX{} ring circuit, with a logarithmic time axis.
    Expval denotes statevector expectation values, and Noise denotes noisy density matrices.
    \jaqsi integrates the matrix equation for all $21n$ pulse evolutions instead of using its closed-form pulse solve.
    Each point is the mean time per batch of \num{2} over \num{20} iterations.
    Error bars denote the standard deviation.}
  \label{fig:benchmarks-absolute-pulse-ode}
\end{figure}

\autoref{fig:benchmarks-absolute-pulse-ode} shows that the matrix-ODE configuration is at most \num{1.5} times slower than the closed-form baseline of \autoref{fig:benchmarks-absolute-pulse}.
In the plotted modes, \jaqsi remains \numrange{5.7}{8.9} times faster than PennyLane, \numrange{13}{277} times faster than QuTiP and \numrange{5.1}{13} times faster than dynamiqs (\autoref{tab:speedup}).
Under noise, the two \jaqsi configurations have similar runtimes at larger qubit counts, where density-matrix operations dominate.
The fixed-step commutator-free Magnus integrators of \autoref{sec:arch_pulse} are not exercised in this benchmark.

\end{document}

%% file: acronyms.tex
\setabbreviationstyle[acronym]{long-short}
\glssetcategoryattribute{acronym}{nohyperfirst}{true}

\newacronym{fcc}{FCC}{Fourier coefficient correlation}
\newacronym{fft}{FFT}{fast Fourier transform}
\newacronym{fm}{FM}{feature map}
\newacronym{ftqc}{FTQC}{fault‑tolerant quantum computing}
\newacronym{hea}{HEA}{hardware‑efficient ansatz}
\newacronym{hep}{HEP}{high‑energy physics}
\newacronym{kl}{KL}{Kullback‑Leibler}
\newacronym{lhc}{LHC}{Large Hadron Collider}
\newacronym{lo}{LO}{leading order}
\newacronym{ml}{ML}{machine learning}
\newacronym{mlp}{MLP}{multi‑layer perceptron}
\newacronym{mse}{MSE}{mean squared error}
\newacronym{nisq}{NISQ}{noisy intermediate‑scale quantum}
\newacronym{nlo}{NLO}{next‑to‑leading order}
\newacronym{pca}{PCA}{principal component analysis}
\newacronym{pcc}{PCC}{pearson correlation coefficient}
\newacronym{pqc}{PQC}{parameterised quantum circuit}
\newacronym{qaoa}{QAOA}{quantum approximate optimisation algorithm}
\newacronym{qc}{QC}{quantum computing}
\newacronym{qcd}{QCD}{quantum chromodynamics}
\newacronym{qfm}{QFM}{quantum Fourier model}
\newacronym{qk}{QK}{quantum kernel}
\newacronym{qml}{QML}{quantum machine learning}
\newacronym{qnn}{QNN}{quantum neural network}
\newacronym{rff}{RFF}{random Fourier features}
\newacronym{spam}{SPAM}{state preparation and measurement}
\newacronym{qoc}{QOC}{quantum optimal control}
\newacronym{rwa}{RWA}{rotating wave approximation}
\newacronym{drag}{DRAG}{derivative removal by adiabatic gate}
\newacronym{dp}{DP}{Dormand--Prince}
\newacronym{jit}{JIT}{just-in-time}
\newacronym{xla}{XLA}{Accelerated Linear Algebra}

%% file: commands.tex
\newcommand{\eg}{\emph{e.g.}\xspace}
\newcommand{\ie}{\emph{i.e.}\xspace}

\newcommand{\jaqsi}{JAQSI\xspace}
\newcommand{\imag}{\mathrm{i}}

\newcommand{\X}{\ensuremath{\mathrm{X}}}
\newcommand{\Y}{\ensuremath{\mathrm{Y}}}
\newcommand{\Z}{\ensuremath{\mathrm{Z}}}
\newcommand{\RX}{\ensuremath{\mathrm{R}}_{X}}
\newcommand{\RY}{\ensuremath{\mathrm{R}}_{Y}}
\newcommand{\RZ}{\ensuremath{\mathrm{R}}_{Z}}
\newcommand{\CX}{\ensuremath{\mathrm{C}\X}}
\newcommand{\CZ}{\ensuremath{\mathrm{C}\Z}}
\newcommand{\CRX}{\ensuremath{\mathrm{C}\RX}}
\newcommand{\hadamard}{\ensuremath{\mathrm{H}}}
\newcommand{\hamiltonian}{\hat{H}}

\newcommand{\cmark}{\ding{51}}%
\newcommand{\xmark}{\ding{55}}%

\definecolor{color1}{HTML}{009682}
\definecolor{color2}{HTML}{DF9B1B}
\definecolor{color3}{HTML}{23A1E0}
\definecolor{color4}{HTML}{002D4C}

\ifbool{anonymous}{}{\renewcommand{\blackout}[1]{#1}}
\ifbool{anonymous}{\newcommand{\genpartemail}[2]{\href{mailto:xxx.xxx@xx.xx}{\blackout{#1}}}}{\newcommand{\genpartemail}[2]{\href{mailto:#1@#2}{#1}}}
\ifbool{anonymous}{\newcommand{\geneurl}[1]{\href{https://thiscatexists.com/}{\blackout{#1}}}}{\newcommand{\geneurl}[1]{\url{#1}}}
\ifbool{anonymous}{\renewcommand{\orcidlink}[1]{}}{}

%% file: figures/jaqsi_overview.tex
\providecommand{\jqmod}[2]{#1\\[1pt]{\ttfamily\footnotesize\textcolor{black!55}{#2}}}
\pgfmathsetmacro{\jqpad}{0.08}
\providecommand{\jqarrowgap}{0.5pt}        
\pgfmathsetmacro{\jqboxh}{0.66+2*\jqpad}   
\pgfmathsetmacro{\jqdrop}{(1.53-\jqboxh)/2}  
\pgfmathsetmacro{\jqdropfar}{2.295-\jqboxh/2}
\begin{tikzpicture}[
    font=\small,
    module/.style={
        draw=#1, fill=#1!8, rounded corners=2pt, semithick,
        minimum width=2.7cm, minimum height=\jqboxh cm, inner ysep=\jqpad cm,
        align=center, text=black,
    },
    cluster/.style={
        draw=#1, fill=#1!4, rounded corners=6pt, line width=1pt,
        inner sep=0.3cm,
    },
    ctitle/.style={font=\small\bfseries, text=color4},
    flow/.style={-{Stealth[length=2mm]}, line width=0.7pt,
                 draw=color4!75, rounded corners=3pt,
                 shorten >=\jqarrowgap, shorten <=\jqarrowgap},
    flowjoin/.style={line width=0.7pt, draw=color4!75, rounded corners=3pt,
                     shorten <=\jqarrowgap},
]

\coordinate (bL) at (-1.35, 0);
\coordinate (bR) at (10.95, 0);

\node[module=color3] (script) at (1.6, 0) {\jqmod{Script}{script.py}};
\node[module=color3] (gates)  at (8.0, 0) {\jqmod{Gates}{gates.py}};

\node[module=color1] (tape)  at (0,   -1.53) {\jqmod{Tape}{tape.py}};
\node[module=color1] (mem)   at (3.2, -1.53) {\jqmod{Memory}{memory.py}};
\node[module=color1] (unit)  at (6.4, -1.53) {\jqmod{Unitary Gates}{unitary.py}};
\node[module=color1] (pulse) at (9.6, -1.53) {\jqmod{Pulse Gates}{pulses.py}};

\node[module=color2] (sim)     at (0,   -3.06) {\jqmod{Simulation}{simulation.py}};
\node[module=color2] (gateset) at (3.2, -3.06) {\jqmod{Gateset}{gateset.py}};
\node[module=color2] (noise)   at (6.4, -3.06) {\jqmod{Noise}{noise.py}};
\node[module=color2] (evo)     at (9.6, -3.06) {\jqmod{Evolution}{evolution.py}};
\node[module=color2, minimum width=12.3cm] (ops) at (4.8, -4.11)
    {\jqmod{Operations}{operations.py}};

\node[module=color4] (draw)   at (13.5, 0)     {\jqmod{Drawing}{drawing.py}};
\node[module=color4] (qoc)    at (13.5, -1.53) {\jqmod{QOC}{qoc.py}};
\node[module=color4] (math)   at (13.5, -3.06)  {\jqmod{Math}{math.py}};
\node[module=color4] (paulis) at (13.5, -4.11) {\jqmod{Paulis}{paulis.py}};


\node[ctitle, anchor=south west] at ($(script.north -| tape.north west)+(-0.8,0.05)$) {Front End};
\node[ctitle, anchor=south west] at ($(tape.north west)+(-0.8,0.05)$)               {Orchestration};
\node[ctitle, anchor=south west] at ($(sim.north west)+(-0.8,0.05)$)                {Engine};
\node[ctitle, anchor=south east] at ($(draw.north east)+(0.8,0.05)$)               {Helpers};

\draw[flow] ($(script.south)+(-0.55,0)$) -- ++(0,-\jqdrop) -| ($(tape.north)+(0.55,0)$);
\draw[flow] ($(script.south)+( 0.55,0)$) -- ++(0,-\jqdrop) -| (mem.north);
\draw[flow] ($(gates.south)+(-0.55,0)$)  -- ++(0,-\jqdrop) -| (unit.north);
\draw[flow] ($(gates.south)+( 0.55,0)$)  -- ++(0,-\jqdrop) -| (pulse.north);
\draw[flow] (script.south) -- ++(0,-\jqdropfar) -| ($(sim.north)+(0.55,0)$);
\draw[flow] ($(unit.south)+(-0.55,0)$) -- ++(0,-\jqdrop) -| (gateset.north);
\draw[flow] (unit.south)  -- (noise.north);
\draw[flow] (pulse.south) -- (evo.north);
\draw[flow] (ops.west) -- ++(-0.95,0) |- (tape.west);
\draw[flow] ($(script.north)+(0,0)$) -- ++(0,0.4) -| (draw.north);
\draw[flow] (qoc.west)  -- ++(-0.6,0) |- ($(gates.east)+(0,0)$);
\draw[flowjoin] (math.west) -- ++(-0.45,0) |- ($(ops.east)+(0.15,0)$);
\draw[flow] (paulis.west) -- (ops.east);

\end{tikzpicture}

%% file: figures/benchmark_circuit.tex
\begin{subfigure}[t]{0.53\columnwidth}
    \centering
    \scalebox{0.75}{%
    \tikzset{operator/.append style={rounded corners=2pt}}%
    \begin{quantikz}[column sep=4pt, row sep={20pt,between origins}]
        \gate[][1.5pt][1.8em]{\RX(x)} & \gate[][1.5pt][1.8em]{\RZ(w)} & \gate[][1.5pt][1.8em]{\RX(w)} & \ctrl{1} &          & \targ{}   & \\
        \gate[][1.5pt][1.8em]{\RX(x)} & \gate[][1.5pt][1.8em]{\RZ(w)} & \gate[][1.5pt][1.8em]{\RX(w)} & \targ{}  & \ctrl{1} &           & \\
        \gate[][1.5pt][1.8em]{\RX(x)} & \gate[][1.5pt][1.8em]{\RZ(w)} & \gate[][1.5pt][1.8em]{\RX(w)} &          & \targ{}  & \ctrl{-2} &
    \end{quantikz}%
    }
    \caption{Gate level.}
    \label{fig:benchmark_circuit_gate}
\end{subfigure}%
\hfill
\begin{subfigure}[t]{0.46\columnwidth}
    \centering
    \scalebox{0.75}{%
    \tikzset{operator/.append style={rounded corners=2pt}}%
    \begin{quantikz}[column sep=4pt, row sep={20pt,between origins}]
        \gate[][1.5pt][1.8em]{\hadamard} & \ctrl{1}                         &                                  & \gate[][1.5pt][1.8em]{\RX(\phi)} & \\
        \gate[][1.5pt][1.8em]{\hadamard} & \gate[][1.5pt][1.8em]{\RX(\phi)} & \ctrl{1}                         &                                  & \\
        \gate[][1.5pt][1.8em]{\hadamard} &                                  & \gate[][1.5pt][1.8em]{\RX(\phi)} & \ctrl{-2}                        &
    \end{quantikz}%
    }
    \caption{Pulse level.}
    \label{fig:benchmark_circuit_pulse}
\end{subfigure}

\vspace{4pt}

\begin{subfigure}{\columnwidth}
    \centering
    \scalebox{0.8}{%
    \tikzset{operator/.append style={rounded corners=2pt}}%
    \begin{quantikz}[column sep=6pt, row sep={20pt,between origins}]
        \lstick{$c$} &                                &                                 & \ctrl{1} &                                  & \ctrl{1} &                                 & \\
        \lstick{$t$} & \gate[][1.5pt][1.8em]{\RZ(\pi / 2)} & \gate[][1.5pt][1.8em]{\RY(\phi / 2)} & \targ{}  & \gate[][1.5pt][1.8em]{\RY(-\phi / 2)} & \targ{}  & \gate[][1.5pt][1.8em]{\RZ(-\pi / 2)} &
    \end{quantikz}%
    }
    \caption{Decomposition of a single \CRX{} gate.}
    \label{fig:benchmark_circuit_crx}
\end{subfigure}

%% file: figures/speedup.tex
\sisetup{range-phrase={\text{--}}}
\begin{tblr}{colspec={lcccc}, row{1}={font=\bfseries}}
  \toprule
  Simulator & Expval & Density & Noise & Grad \\
  \midrule
  \SetCell[c=5]{l} \emph{Gate level} & & & & \\
  PennyLane & \numrange{0.88}{2.6} & \numrange{0.78}{0.99} & \numrange{0.97}{5.9} & \numrange{0.97}{2.5} \\
  PennyLane Lightning & \numrange{4.4}{8.4} & \numrange{3.3}{12} & -- & \numrange{5.5}{33} \\
  Qiskit & \numrange{4.7}{7.0} & \numrange{4.8}{68} & \numrange{1.5}{12} & -- \\
  Qibo & \numrange{6.5}{13} & \numrange{7.4}{49} & \numrange{3.2}{23} & -- \\
  Qulacs & \numrange{1.7}{2.0} & \numrange{1.9}{43} & \numrange{3.0}{13} & \numrange{2.1}{3.3} \\
  \midrule
  \SetCell[c=5]{l} \emph{Pulse level} & & & & \\
  PennyLane & \numrange{9.1}{12} & -- & \numrange{6.1}{9.3} & -- \\
  QuTiP & \numrange{158}{375} & -- & \numrange{14}{231} & -- \\
  dynamiqs & \numrange{7.5}{17} & -- & \numrange{7.2}{13} & -- \\
  \midrule
  \SetCell[c=5]{l} \emph{Pulse level (matrix ODE)} & & & & \\
  PennyLane & \numrange{6.9}{8.9} & -- & \numrange{5.7}{8.4} & -- \\
  QuTiP & \numrange{110}{277} & -- & \numrange{13}{175} & -- \\
  dynamiqs & \numrange{5.5}{13} & -- & \numrange{5.1}{13} & -- \\
  \bottomrule
\end{tblr}